\documentclass[a4paper]{article}

\usepackage[english]{babel}
\usepackage[utf8x]{inputenc}
\usepackage[T1]{fontenc}

\usepackage[a4paper,top=3cm,bottom=2cm,left=3cm,right=3cm,marginparwidth=1.75cm]{geometry}

\usepackage{setspace}
\usepackage{amssymb}

\usepackage{cite}

\usepackage{amsmath}
\usepackage{graphicx}
\usepackage[colorinlistoftodos]{todonotes}
\usepackage[colorlinks=true, allcolors=blue]{hyperref}

\title{Dynamically reversible and strong circular dichroism based on Babinet-invertible chiral metasurfaces}

\author{Xiaoqing Luo$^{1,2}$, Fangrong Hu$^{3}$, and Guangyuan Li$^{1,2,4,*}$}

\date{}
\begin{document}
\maketitle

\begin{spacing}{2.0}

\noindent \large$^1$CAS Key Laboratory of Human-Machine Intelligence-Synergy Systems, Shenzhen Institutes of Advanced Technology, Chinese Academy of Sciences, Shenzhen, 518055 China

\noindent  $^2$Guangdong-Hong Kong-Macao Joint Laboratory of Human-Machine Intelligence-Synergy Systems, Chinese Academy of Sciences, Shenzhen Institutes of Advanced Technology, Shenzhen 518055, China

\noindent  $^3$Guangxi Key Laboratory of Optoelectronic Information Processing, Guilin University of Electronic
Technology, Guilin 541004, China

\noindent  $^4$Shenzhen College of Advanced Technology, University of Chinese Academy of Sciences, Shenzhen 518055, China


\noindent *Corresponding author: gy.li@siat.ac.cn

\end{spacing}

\begin{abstract}
We propose a Babinet-invertible chiral metasurface for achieving dynamically reversible and strong circular dichroism (CD). The proposed metasurface is composed of VO$_2$-metal hybrid structure, and when VO$_2$ transits between the dielectric state and the metallic state, the metasurface unit cell switches between complementary structures that are designed according to the Babinet principle. This leads to a large and reversible CD tuning range between $\pm 0.5$ at 0.97~THz, which is larger than the literature. We attribute the CD effect to extrinsic chirality of the proposed metasurface. We envision that the Babinet-invertible chiral metasurface proposed here will advance the engineering of active and tunable chiro-optical devices and promote their applications.
\end{abstract}

\newpage

Manipulation of optical chirality has a diverse range of applications, such as polarization conversion \cite{2019Spin}, bio-sensing \cite{2019Plasmonic}, polarization-sensitive spectrometry \cite{2019THz,2019Terahertz}, and telecommunications \cite{2017Scenario}. Chiral metamaterial emerges as a new and exciting platform for providing remarkable chiro-optical effects including circular dichroism \cite{oh2015chiral}, optical activity \cite{xie2020lattice} and asymmetric transmission \cite{huang2017tunable}. Among these, circular dichroism (CD) manifests itself as a difference between the transmittances for the left- and right-handed circularly polarized  (LCP and RCP) waves, which can be calculated with 
\begin{equation}
\label{eq:CD}
CD= T_{\rm LCP}-T_{\rm RCP}\,,
\end{equation}
where $T_{\rm LCP}$ and $T_{\rm RCP}$ are the transmittances of the LCP and the RCP waves, respectively. Making use of circular dichroism, broadband circular polarizers \cite{zhao2012twisted}  and chiral beam splitters \cite{turner2013miniature} have been demonstrated. 


Over the years, interest has shifted from how to increase the chiral optical response into how to realize switchable chirality under a certain external stimulus \cite{hentschel2017chiral}. In order to achieve dynamic tuning of CD, one approach is to reconfigurate the structural geometry \cite{neubrech2020reconfigurable}. This is usually achieved by using reconfigurable 3D plasmonic metamolecules. For example, Kuzyk {\sl et al.} \cite{kuzyk2014reconfigurable} proposed reconfigurable three-dimensional plasmonic metamolecules and demonstrated reversible CD through DNA-regulated conformational changes at the nanoscale. Lan {\sl et al.} \cite{lan2018dna} also demonstrated dynamical tuning of CD through DNA-guided structural reconfiguration of self-assembled gold nanorod helix superstructures.
To remove the need for geometrical reconfiguration, in 2012 Zhang {\sl et al.} \cite{zhang2012photoinduced} demonstrated optically controlled CD switching between $-0.3$ and $0.3$ in a 3D chiral terahertz metamaterial with embedded photo-active silicon. Since then metamaterials incorporating active materials, such as silicon, graphene and phase change materials, have been the focus for dynamical tuning of CD effect. For example, Kenannakis {\sl et al.} \cite{kenanakis2014optically} proposed a bi-layer uniaxial chiral metamaterial incorporating photoconducting Si and showed that CD between 0 and 0.16 can be dynamically tuned under external optical pumping. Yin {\sl et al.} \cite{yin2015active} demonstrated reversal of the circular dichroism sign (between $-0.1$ and 0.2) in layered plasmonic chiral metamaterial incorporating phase change material Ge$_3$Sb$_2$Te$_6$. Lv {\sl et al.} \cite{lv2016hybrid} proposed two 90$^\circ$ twisted E-shaped resonators with embedded vanadium dioxide (VO$_2$) and achieved a complete switching effect of CD with a maximum change of 0.45 at around 1.36~THz. Wang {\sl et al.} \cite{wang2018active} also proposed bi-layer hybrid metamaterials integrated with vanadium dioxide (VO2) showing switchable CD between 0 and 0.4. However, these 3D chiral structures are bulky, and difficult to fabricate and integrate.



Since the first experimental demonstration by Papakostas {\sl et al.} \cite{papakostas2003optical}, planar metasurfaces, which are much easier to fabricate and integrate compared with the above 3D chiral metamaterials, and which exhibit extrinsic chirality under oblique incidence, have been a promising platform for achieving large CD \cite{plum2009extrinsic,yoo2019metamaterials,2020Emerging,intaravanne2020recent}. By further incorporating active materials such as graphene, Huang {\sl et al.} \cite{huang2018graphene} proposed a graphene–metal hybrid chiral metasurface and achieved reversible CD between $-0.05$ and 0.04 in the mid-infrared regime by varying the Fermi energy of graphene. Zhou {\sl et al.} \cite{zhou2019tunable} also proposed a graphene metasurface and showed reversible and large CD between $-0.25$ and 0.4. However, these CD tuning ranges are relatively small, hindering the potential applications.





In this letter, we tackle this challenge by proposing a Babinet-invertible chiral metasurface for achieving dynamically switchable and large circular dichroism chiral metasurface based on Babinet's principle. We will show that by changing the phase of VO$_2$ between the dielectric state and the metallic state, which can be done thermally, optically or electrically \cite{NPGAM2018_VO2Rev}, the CD can be tuned between 0.5 and $-0.5$ at 0.97~THz. This CD tuning range is larger compared with the above 3D metamaterials \cite{zhang2012photoinduced,kenanakis2014optically,yin2015active,lv2016hybrid,wang2018active} and planar metasurfaces \cite{huang2018graphene,zhou2019tunable}. The operation principle will be clarified based on the Baninet's principle and extrinsic chirality. The effects of the VO$_2$ conductivity on the CD tuning performance will also be investigated. 

\begin{figure}[htbp]
\centering
\includegraphics[width=8.4cm]{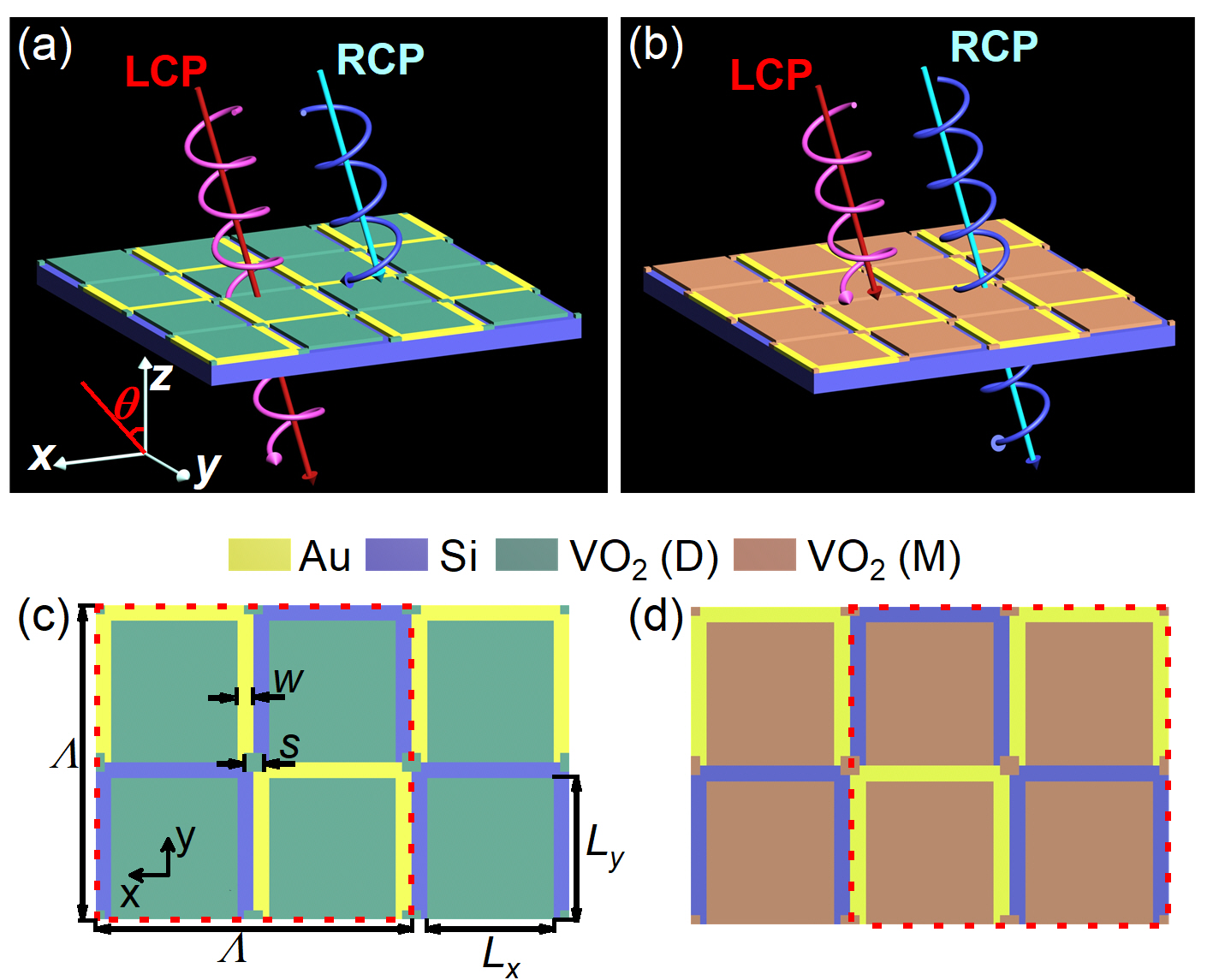}
\caption{Schematics of the proposed Babinet-invertible chiral metasurface for achieving reversible and strong CD effect. (a) When VO$_2$ is in the dielectric state (in green, denoted as ``VO$_2$(D)''), LCP terahertz wave transmits through the metasurface while RCP wave is blocked. (b) When VO$_2$ is in the metallic state (in brown, denoted as ``VO$_2$(M)''), LCP wave transmits through the metasurface while RCP wave is blocked. (c)(d) Top views of 
the metasurface unit cells (in dashed boxes) for VO$_2$(D) and VO$_2$(M), respectively. $\Lambda=100~\mu$m, $w=7~\mu$m, $L_{x}=L_{y}=36~\mu$m, $s=8~\mu$m.}
\label{fig:schem}
\end{figure}

Figure~\ref{fig:schem} illustrates the proposed Barbinet-invertible chiral metasurface composed of hybrid gold-VO$_2$ structures on top of a high-resistivity silicon substrate. When VO$_2$ is in the dielectric state, denoted as ``VO$_2$(D)'' and in green, the interactions between the VO$_2$ structures and the incident terahertz wave are negligible compared with those between the gold structures and the terahertz wave, thus the  metasurface unit cell is effectively composed of the left-top and right-bottom U-shaped gold ridges, as illustrated in the red dashed box in Fig.~\ref{fig:schem}(c). In this scenario, LCP terahertz wave can transmit through the metasurface whereas RCP wave is blocked, suggesting strong circular dichroism effect, as illustrated by Fig.~\ref{fig:schem}(a). 

When VO$_2$ is in the metallic state, denoted as ``VO$_2$(M)'' and in brown, VO$_2$ behaves like gold. If we approximately treat metallic VO$_2$ as gold and shift the unit cell in Fig.~\ref{fig:schem}(c) by half a period, the equivalent metasurface unit cell is composed of the left-top and right-bottom U-shaped slots in a gold film, as illustrated in the red dashed box in Fig.~\ref{fig:schem}(d). Comparing the effective unit cells in Figs.~\ref{fig:schem}(c)(d), we find these are exactly complementary structures. According to the Babinet's principle \cite{falcone2004babinet}, RCP terahertz wave will transmit through the metasurface whereas LCP wave will be blocked in this scenario, as illustrated by Fig.~\ref{fig:schem}(b). Therefore, taking advantage of the Babinet’s principle, the proposed Babinet-invertible chiral metasurface can exhibit reversible circular dichroism effect. 



The reversible polarization dependent transmission performance was numerically evaluated using the frequency-domain solver in CST Microwave Studio. Frequency-dependent transmission is obtained as $t(f)=S_{12}(f)$, and the corresponding transmittance is then calculated with $T=|t|^2$. In the simulations, we take the periods of the metasurface unit cell in both $x$ and $y$ directions to be $\Lambda= 100~\mu$m, and take the thicknesses of gold and VO$_2$ be 200~nm and 400~nm, respectively. Unless otherwise specified, the incident angle is set to be $\theta=60^\circ$. We adopted frequency-dependent permittivities of lossy gold and loss-free silica substrate from the build-in material library. Drude model was adopted to describe the frequency-dependent permittivities of VO$_{2}$ in the terahertz regime \cite{zhu2012effect},
\begin{equation}
    \label{eq:Epsilon}
    \varepsilon(\omega)=\varepsilon_{\infty}-\frac{\omega_{\rm p}^{2}(\sigma)}{\omega^{2}+i\gamma\omega}\,
\end{equation}
where $\varepsilon_{\infty}$ is the permittivity at high frequency limit, $\omega_{p}(\sigma)$ is the conductivity-dependent plasmon frequency, $\sigma$ is the conductivity, and $\gamma=5.75\times 10^{13}$~rad/s is the collision frequency assumed to be independent of the conductivity. A further examination reveals that $\omega_{p}^{2}(\sigma)$ can be approximately expressed as $\omega^{2}_{\rm p}(\sigma)=(\sigma/\sigma_0)\cdot\omega^{2}_{\rm p}(\sigma_{0})$ with $\omega_{\rm p}(\sigma_0)= 1.4\times10^{15}$~rad/s for $\sigma_{0}=3\times10^5$~S/m \cite{zhu2012effect}. For VO$_2$ in the dielectric and the metallic states, we take $\sigma$ to be $40$~S/m and $4\times10^5$~S/m, respectively.

\begin{figure}[htbp]
\centering
\includegraphics[width=8.8cm]{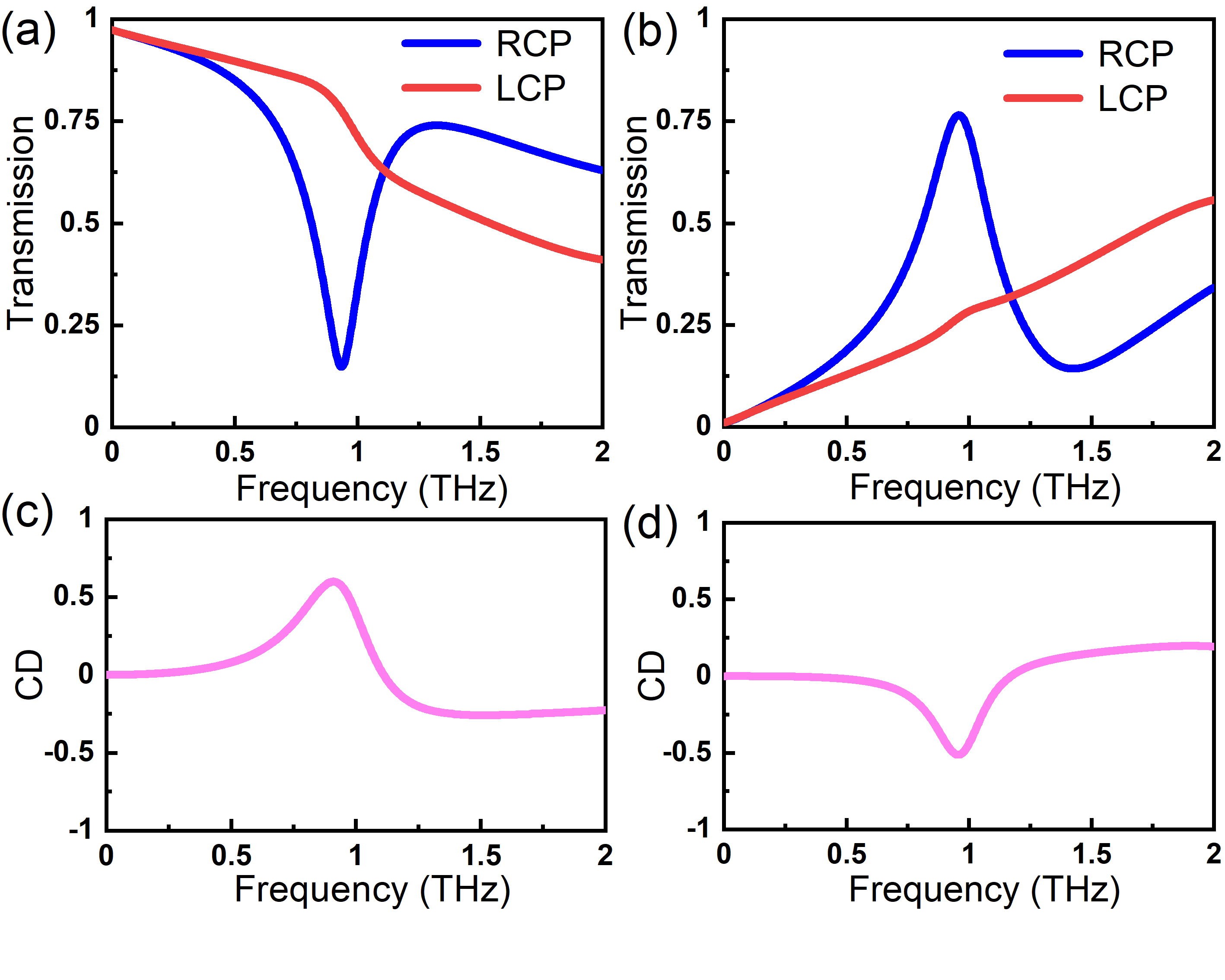}
\caption{(a)(b) Simulated transmission spectra $|t(f)|$ for LCP and RCP waves and (c)(d) the corresponding circular dichorism when VO$_2$ is in the (a)(c) dielectric or (b)(d) metallic state.}
\label{fig:TvsWv}
\end{figure}

Figure~\ref{fig:TvsWv}(a) shows that when VO$_2$ is in the dielectric state, the transmission decreases as the operation frequency increases for both the LCP and the RCP terahertz waves. There exists a pronounced dip locating at frequency of $f=0.94$~THz for the RCP wave. When VO$_2$ is in the metallic state, however, the transmission spectra behave reversely: the transmission increases with the frequency, and there is a transmission peak at $f=0.96$~THz for the RCP wave, as shown in Fig.~\ref{fig:TvsWv}(b). The peak transmission frequency is slightly blue shifted by 2$\%$ compared with the dip one in Fig.~\ref{fig:TvsWv}(a). 


With the transmission spectra for the LCP and the RCP waves, the corresponding CD spectra can be calculated with Eq.~(\ref{eq:CD}) and are depicted in Figs.~\ref{fig:TvsWv}(c)(d). Results show that the maximum value reaches $CD=0.59$ at $f=0.91$~THz for the dielectric VO$_2$ state and $CD=-0.51$ at $f=0.96$~THz for the metallic VO$_2$ state. Strikingly, at 0.97~THz the circular dichroism of the proposed metasurface switches reversely between 0.5 and $-0.5$ when VO$_2$ transits between the dielectric state and the metallic state.

\begin{figure}[htbp]
\centering
\includegraphics[width=8.8cm]{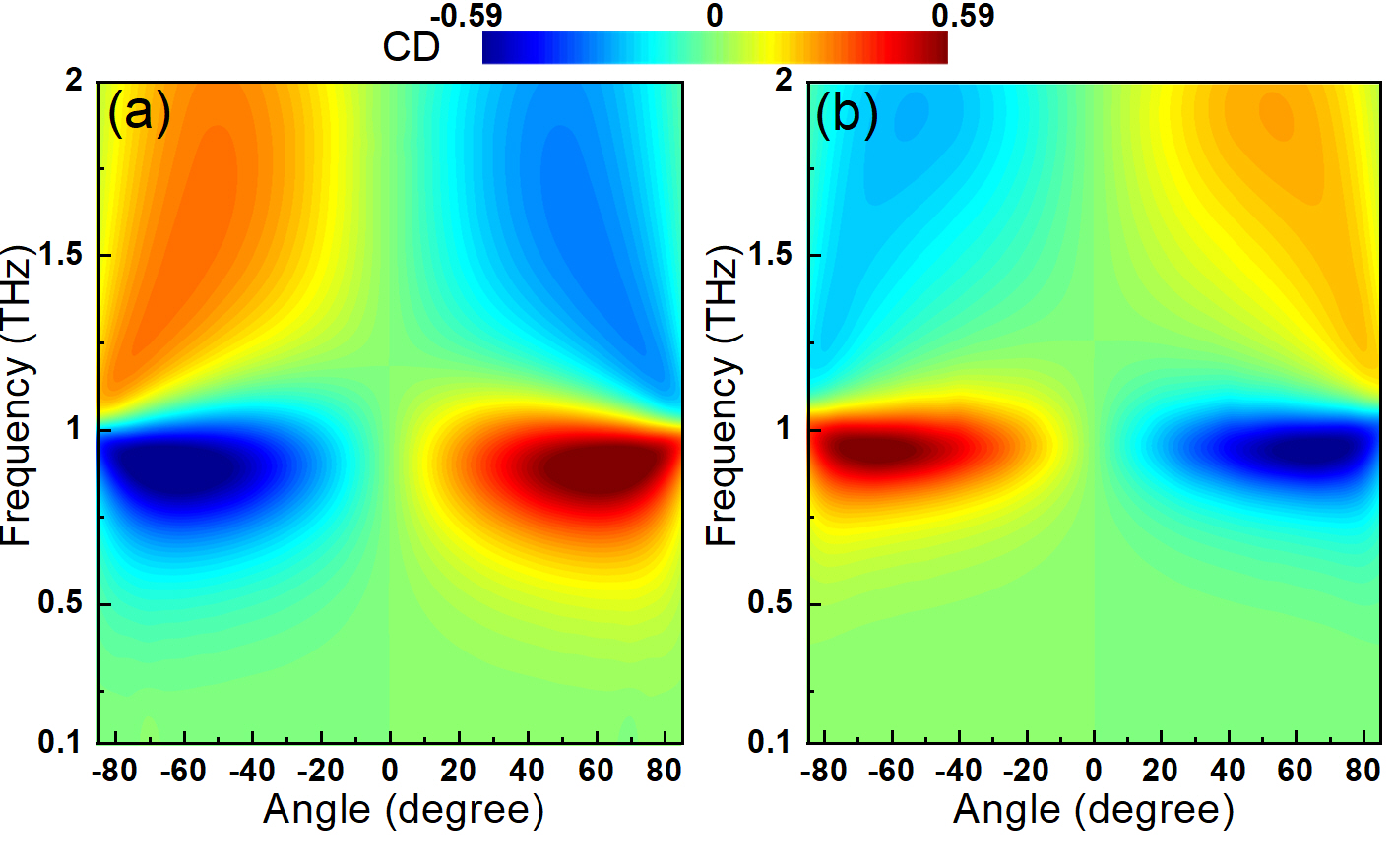}
\caption{CD spectra for various incident angles between $-85^\circ$ and $85^\circ$ for (a) VO$_2$(D) and (b) VO$_2$(M).}
\label{fig:CDvsAng}
\end{figure}

In order to understand the physics underlying the reversible and strong CD effect of the proposed metasurface, we first calculated the dependence of CD spectra on the incident angle and plotted the results in Fig.~\ref{fig:CDvsAng}. We find that under normal incidence ($\theta=0^{\circ}$), the proposed metasurface exhibits no CD effect regardless of the operation frequency and the VO$_2$ state (dielectric or metallic). This suggests that the CD effect of the proposed metasurface is not intrinsic. As the incidence angle increases, the CD effect first becomes pronounced and then gradually decreases for frequencies around $f=0.95$~THz. The CD spectra are anti-symmetric with respect to the $\theta=0^\circ$ axis. Figs.~\ref{fig:CDvsAng}(a)(b) for the dielectric and the metallic VO$_2$ states are slightly different. These differences arise because metallic VO$_2$ with $\sigma=4 \times 10^5 $~S/m cannot behave the same as gold, and dielectric VO$_2$ as well as the substrate also affects the performances. The maximum CD values are $\pm0.59$ at $f=0.91$~THz and $\theta=\pm 60^\circ$ for the dielectric VO$_2$ state, and $\pm0.51$ at $f=0.96$~THz and $\theta=\pm 60^\circ$ for the metallic VO$_2$ state. The strong angular dependence indicates that the CD effect in the proposed metasurface belongs to extrinsic chirality.

\begin{figure}[htbp]
\centering
\includegraphics[width=8.8cm]{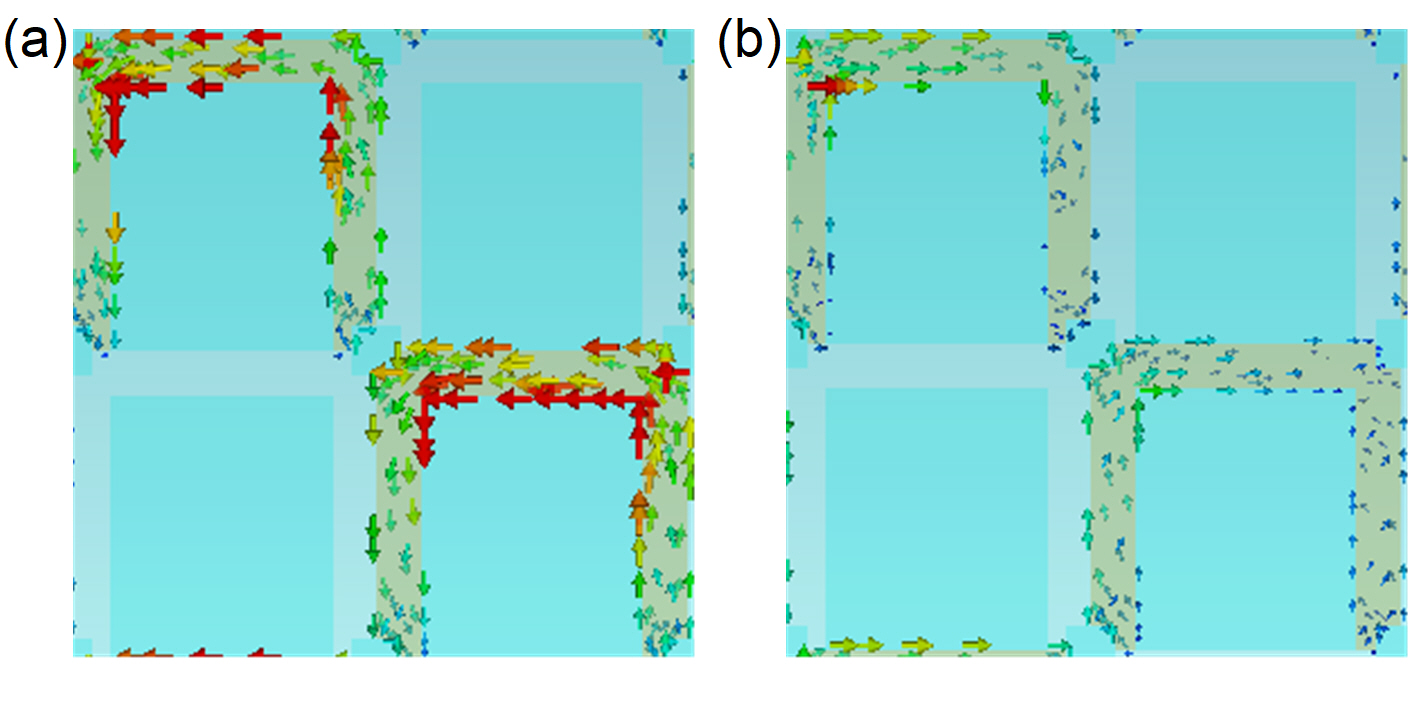}
\caption{Current distributions in the metasurface unit cell with VO$_2$(D) for (a) RCP and (b) LCP waves. The calculations were performed with $\theta=60^\circ$ and $f=0.94$~THz.}
\label{fig:Current}
\end{figure}

To further understand the distinct transmission and strong CD effect for the LCP and the RCP waves with oblique incidence of $\theta=60^\circ$, in Fig.~\ref{fig:Current} we plot the current distributions of the proposed metasurface with dielectric VO$_2$ at $f=0.94$~THz. Results show that for the RCP wave, strong counterclockwise currents are excited in the U-shaped VO$_2$ structures, whereas for the LCP wave, weak clockwise currents are excited. The strong resonance for the RCP wave results in the transmission dip, whereas the weak one for the LCP wave leads to high transmission. These explain the large difference in transmission in Fig.~\ref{fig:TvsWv}(a) and the large CD effect in Fig.~\ref{fig:TvsWv}(c). The transmission difference in Fig.~\ref{fig:TvsWv}(b) and the large but reverse CD effect in Fig.~\ref{fig:TvsWv}(d) can also be explained similarly.

\begin{figure}[htbp]
\centering
\includegraphics[width=8.8cm]{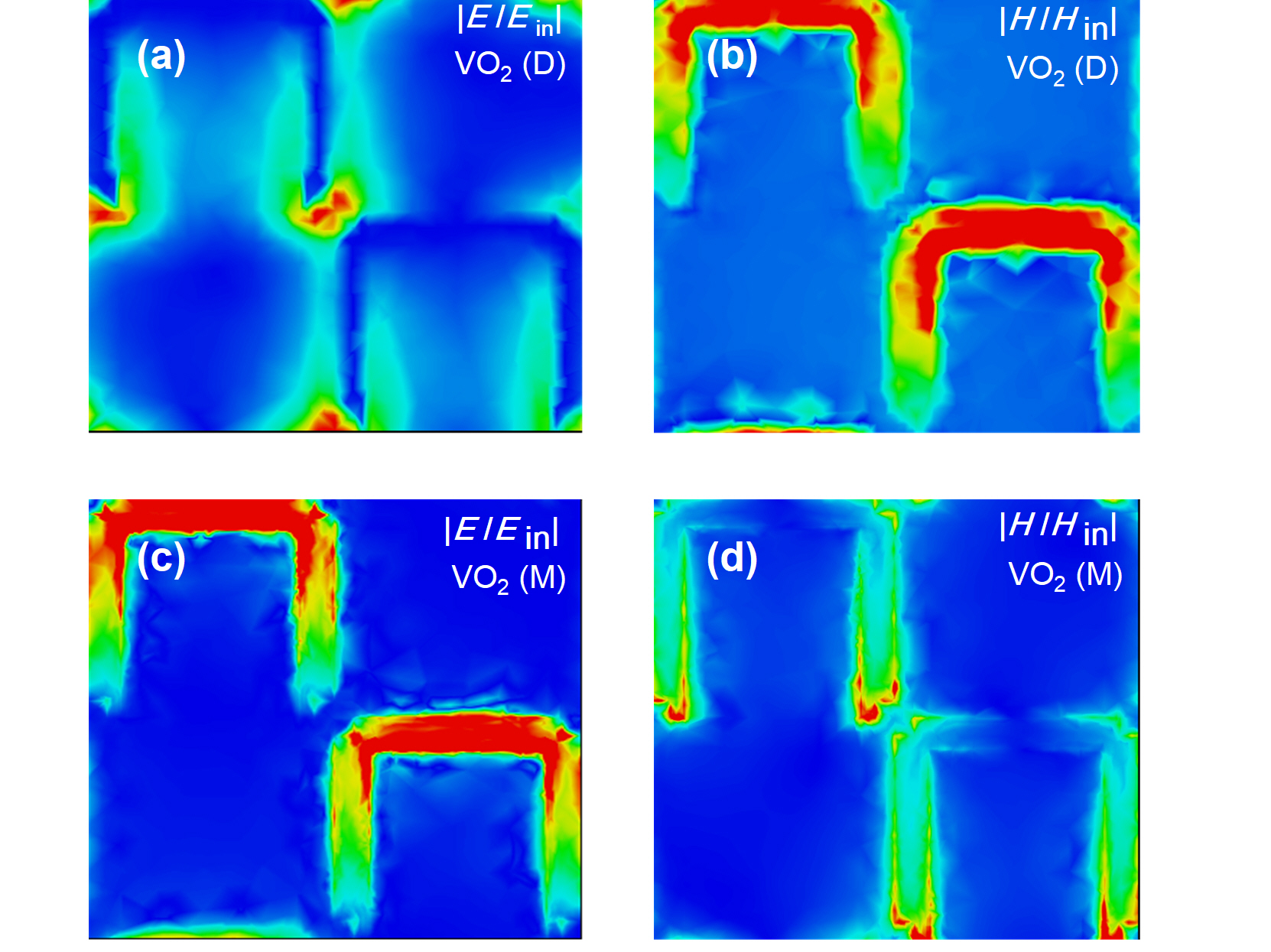}
\caption{(a)(c) Electric field and (b)(d) magnetic field distributions in the metasurface unit cells under obliquely incident RCP wave. The calculations were performed with  $\theta=60^\circ$, and (a)(b) $f=0.94$~THz for VO$_2$(D) and (c)(d) $f=0.96$~THz for VO$_2$(M).}
\label{fig:EHfields}
\end{figure}

The reversible CD effect in the proposed Babinet-invertible chiral metasurface can be understood by the electric and magnetic field distributions for the resonant frequencies of the LCP and the RCP waves when VO$_2$ transits between the dielectric and the metallic state. Figs.~\ref{fig:EHfields}(a)(b) show that electric and magnetic fields are confined to the U-shaped gold structures when VO$_2$ is in the dielectric state. However, when VO$_2$ is in the metallic state, electric and magnetic fields are confined to the U-shaped slots, as shown by Figs.\ref{fig:EHfields}(c)(d). By comparing Figs.~\ref{fig:EHfields}(a)(d) [or (b)(c)], we find the electric (or magnetic) field for dielectric VO$_2$ has similar distributions as the magnetic (or electric) field for metallic VO$_2$. These similarities are consistent with complementary structures designed according to the Babinet principle in the literature \cite{bitzer2011terahertz}.

\begin{figure}[htbp]
\centering
\includegraphics[width=8.8cm]{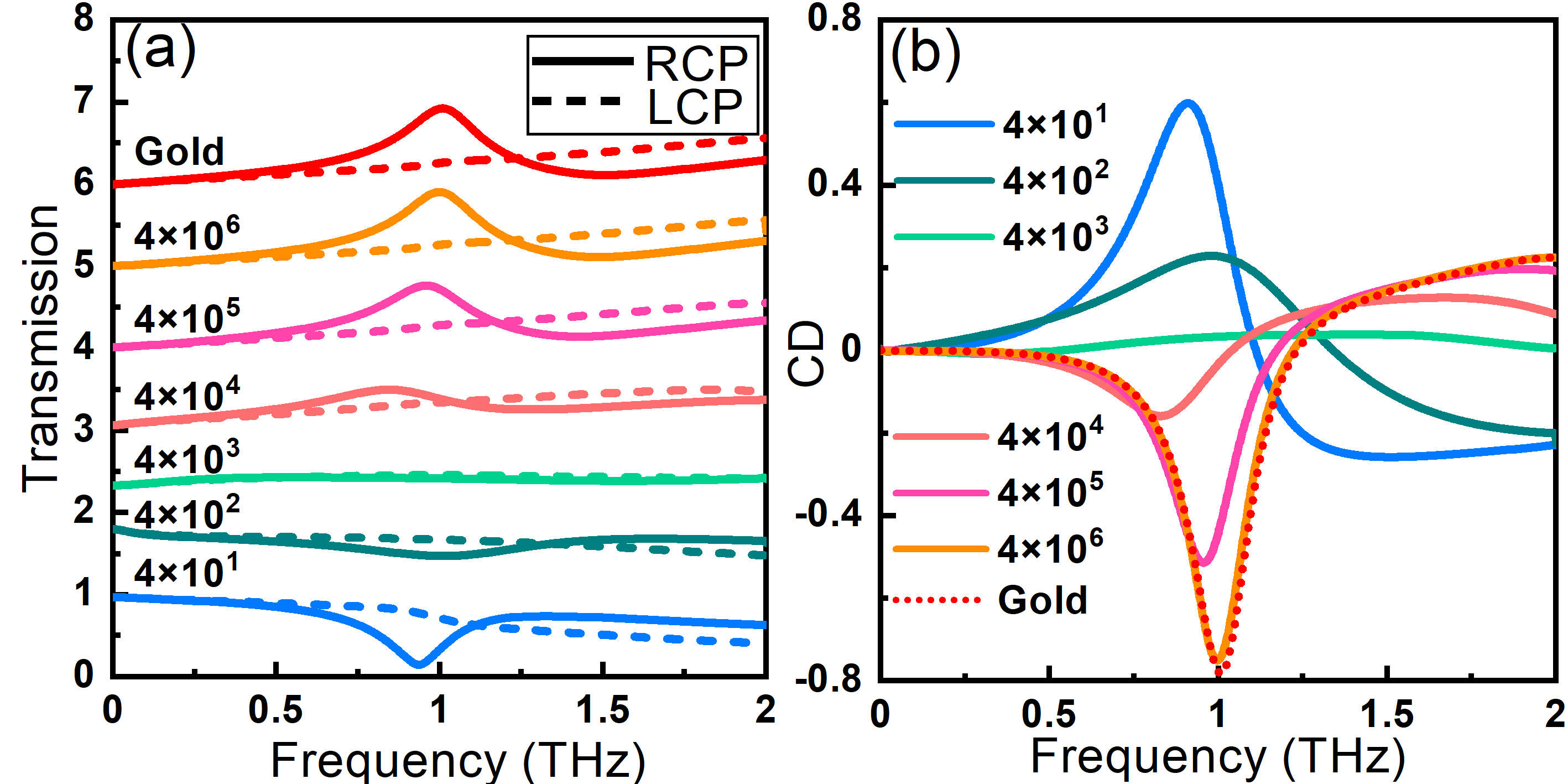}
\caption{(a) Transmission spectra and (b) CD spectra for various conductivities of VO$_2$, which are taken to vary from 40~S/m to $4\times 10^6$~S/m. As a comparison, we also plot the spectra when VO$_2$ is replaced by gold.}
\label{fig:Conduct}
\end{figure}

Until now we have fixed the VO$_2$ conductivities for the dielectric and the metallic states, in experiments, however, the conductivities of the as-fabricated VO$_2$ films may vary significantly \cite{OE2020_VO2AT}. Hereafter we investigate the effects of the VO$_2$ conductivities on the transmission and CD spectra for $\theta = 60^\circ$. Fig.~\ref{fig:Conduct}(a) shows that as the VO$_2$ conductivity increases from $\sigma=40$~S/m to $4\times 10^6$~S/m, the transmission spectra for the RCP first show a dip with decreasing strength and red-shifting frequency, and then exhibit a peak with increasing amplitude and blue-shifting frequency. Interestingly, when $\sigma=4\times 10^3$~S/m, the transmission spectrum for the LCP wave almost overlaps with that for the RCP wave. Correspondingly, the peak CD evolves from 0.59 for $\sigma=40$~S/m, to near zero $\sigma=4\times 10^3$~S/m, and to $-0.75$ for $\sigma=4\times 10^6$~S/m. Therefore, results suggest that in order to achieve strong and reversible CD tuning beyond $\pm 0.5$ for the proposed metasurface, the conductivity difference between the dielectric and metallic VO$_2$ should be three or four orders of magnitude. This requires high quality deposited VO$_2$ films, which were reported by Yun {\sl et al.} \cite{yun2008vanadium} and also by the authors quite recently \cite{OE2020_VO2AT}.

As a comparison, we also calculated the transmission and the CD spectra when VO$_2$ is replaced by gold, the conductivity of which is $\sigma=4.56\times10^7$~S/m in the CST build-in library. We find that these spectra overlap well with those when the conductivity of VO$_2$ is $\sigma=4\times 10^6$~S/m. In other words, if the conductivity of VO$_2$ is above $4\times 10^6$~S/m, it can be effectively treated as gold.



In conclusion, we have proposed a Babinet-invertible chiral metasurface to achieve reversible and strong CD effect. Simulation results have shown that, when VO$_2$ transits between the dielectric state and the metallic state, the transmission spectrum for the RCP wave is flipped, from exhibiting a dip to showing a peak. Correspondingly, the CD switches between 0.5 and $-0.5$ at 0.97~THz. The angular dependent CD spectra have shown that the CD effect of the proposed metasurface is extrinsic. The near-field distributions have revealed that the equivalent metasurface unit cells for dielectric and metallic VO$_2$ are complementary structures satisfying the Babinet principle. By investigating the influences of the VO$_2$ conductivity on the CD tuning performance, we have shown that three or four orders of magnitude large difference between the VO$_2$ conductivities for the dielectric state and the metallic state is required to achieve the large CD tuning range. We expect the proposed Babinet-invertible chiral metasurface will provide a new strategy for realizing active and tunable optical chirality and will find applications in polarization dependent fields.

\section*{Acknowledgments}
This work was supported by Shenzhen Fundamental Research and Discipline Layout project (JCYJ20180507182444250) and China Postdoctoral Science Foundation (2020M682984).

\bibliographystyle{unsrt}
\bibliography{sample}

\begin{thebibliography}{10}

\bibitem{2019Spin}
Shengyan Yang, Zhe Liu, Sha Hu, Ai~Zi Jin, Haifang Yang, Shuang Zhang, Junjie
  Li, and Changzhi Gu.
\newblock Spin-selective transmission in chiral folded metasurfaces.
\newblock {\em Opt. Lett.}, 6(4):3432--3439, 2019.

\bibitem{2019Plasmonic}
Yoon~Young Lee, Ryeong~Myeong Kim, Sang~Won Im, Mani Balamurugan, and Ki~Tae
  Nam.
\newblock Plasmonic metamaterials for chiral sensing applications.
\newblock {\em Nanoscale}, 12:58--66, 2019.

\bibitem{2019THz}
Quan Guo, Yuan Zhang, Zhihui Lyu, Dongwen Zhang, Yindong Huang, Chao Meng,
  Zengxiu Zhao, and Jianmin Yuan.
\newblock {THz} time-domain spectroscopic ellipsometry with simultaneous
  measurements of orthogonal polarizations.
\newblock {\em IEEE Trans. Terahertz Sci. Technol.}, 9(4):422--429, 2019.

\bibitem{2019Terahertz}
Won~Jin Choi, Gong Cheng, Zhengyu Huang, Shuai Zhang, Theodore~B Norris, and
  Nicholas~A Kotov.
\newblock Terahertz circular dichroism spectroscopy of biomaterials enabled by
  kirigami polarization modulators.
\newblock {\em Nat. Mater.}, 18(8):820--826, 2019.

\bibitem{2017Scenario}
Guan Ke, Ai~Bo, Peng Bile, He~Danping, Lin Xue, Wang Longhe, Zhong Zhangdui,
  and Kürner Thomas.
\newblock Scenario modules, ray-tracing simulations and analysis of millimetre
  wave and terahertz channels for smart rail mobility.
\newblock {\em IET Microw.}, 12(4):501--508, 2017.

\bibitem{oh2015chiral}
Sang~Soon Oh and Ortwin Hess.
\newblock Chiral metamaterials: enhancement and control of optical activity and
  circular dichroism.
\newblock {\em Nano Converg.}, 2(24):24, 2015.

\bibitem{xie2020lattice}
Fei Xie, Wei Wu, Mengxin Ren, Wei Cai, and Jingjun Xu.
\newblock Lattice collective interaction engineered optical activity in
  metamaterials.
\newblock {\em Adv. Opt. Mater.}, 8(2):1901435, 2020.

\bibitem{huang2017tunable}
Yuanyuan Huang, Zehan Yao, Fangrong Hu, Changji Liu, Leilei Yu, Yanping Jin,
  and Xinlong Xu.
\newblock Tunable circular polarization conversion and asymmetric transmission
  of planar chiral graphene-metamaterial in terahertz region.
\newblock {\em Carbon}, 119:305--313, 2017.

\bibitem{zhao2012twisted}
Yang Zhao, MA~Belkin, and A~Al{\`u}.
\newblock Twisted optical metamaterials for planarized ultrathin broadband
  circular polarizers.
\newblock {\em Nat. Commun.}, 3(1):870, 2012.

\bibitem{turner2013miniature}
Mark~D Turner, Matthias Saba, Qiming Zhang, Benjamin~P Cumming, Gerd~E
  Schr{\"o}der-Turk, and Min Gu.
\newblock Miniature chiral beamsplitter based on gyroid photonic crystals.
\newblock {\em Nat. Photon.}, 7(10):801--805, 2013.

\bibitem{hentschel2017chiral}
Mario Hentschel, Martin Sch{\"a}ferling, Xiaoyang Duan, Harald Giessen, and
  Na~Liu.
\newblock Chiral plasmonics.
\newblock {\em Sci. Adv.}, 3(5):e1602735, 2017.

\bibitem{neubrech2020reconfigurable}
Frank Neubrech, Mario Hentschel, and Na~Liu.
\newblock Reconfigurable plasmonic chirality: Fundamentals and applications.
\newblock {\em Adv. Mater.}, 32(41):1905640, 2020.

\bibitem{kuzyk2014reconfigurable}
Anton Kuzyk, Robert Schreiber, Hui Zhang, Alexander~O Govorov, Tim Liedl, and
  Na~Liu.
\newblock Reconfigurable 3{D} plasmonic metamolecules.
\newblock {\em Nat. Mater.}, 13(9):862--866, 2014.

\bibitem{lan2018dna}
Xiang Lan, Tianji Liu, Zhiming Wang, Alexander~O Govorov, Hao Yan, and Yan Liu.
\newblock {DNA}-guided plasmonic helix with switchable chirality.
\newblock {\em J. Am. Chem. Soc.}, 140(37):11763--11770, 2018.

\bibitem{zhang2012photoinduced}
Shuang Zhang, Jiangfeng Zhou, Yong~Shik Park, Junsuk Rho, Ranjan Singh,
  Sunghyun Nam, Abul~K Azad, Hou~Tong Chen, Xiaobo Yin, Antoinette~J Taylor,
  and Xiang Zhang.
\newblock Photoinduced handedness switching in terahertz chiral metamolecules.
\newblock {\em Nat. Commun.}, 3(1):1--7, 2012.

\bibitem{kenanakis2014optically}
George Kenanakis, Rongkuo Zhao, N~Katsarakis, M~Kafesaki, CM~Soukoulis, and
  EN~Economou.
\newblock Optically controllable {THz} chiral metamaterials.
\newblock {\em Opt. Express}, 22(10):12149--12159, 2014.

\bibitem{yin2015active}
Xinghui Yin, Martin Sch{\"a}ferling, Ann Katrin~U Michel, Andreas Tittl,
  Matthias Wuttig, Thomas Taubner, and Harald Giessen.
\newblock Active chiral plasmonics.
\newblock {\em Nano. Lett.}, 15(7):4255--4260, 2015.

\bibitem{lv2016hybrid}
TT~Lv, YX~Li, HF~Ma, Z~Zhu, ZP~Li, CY~Guan, JH~Shi, H~Zhang, and TJ~Cui.
\newblock Hybrid metamaterial switching for manipulating chirality based on
  {VO$_2$} phase transition.
\newblock {\em Sci. Rep.}, 6:23186, 2016.

\bibitem{wang2018active}
Sheng~Xiang Wang, Lei Kang, and Douglas~H Werner.
\newblock Active terahertz chiral metamaterials based on phase transition of
  vanadium dioxide ({VO$_2$}).
\newblock {\em Sci. Rep.}, 8(1):189, 2018.

\bibitem{papakostas2003optical}
A~Papakostas, A~Potts, DM~Bagnall, SL~Prosvirnin, HJ~Coles, and NI~Zheludev.
\newblock Optical manifestations of planar chirality.
\newblock {\em Phys. Rev. Lett.}, 90(10):107404, 2003.

\bibitem{plum2009extrinsic}
E~Plum, VA~Fedotov, and NI~Zheludev.
\newblock Extrinsic electromagnetic chirality in metamaterials.
\newblock {\em J. Opt. A: Pure Appl. Opt.}, 11(7):074009, 2009.

\bibitem{yoo2019metamaterials}
SeokJae Yoo and Q~Han Park.
\newblock Metamaterials and chiral sensing: a review of fundamentals and
  applications.
\newblock {\em Nanophotonics}, 8(2):249--261, 2019.

\bibitem{2020Emerging}
Heonyeong Jeong, Younghwan Yang, Hanlyun Cho, Trevon Badloe, Inki Kim, Ren~Min
  Ma, and Junsuk Rho.
\newblock Emerging advanced metasurfaces: Alternatives to conventional bulk
  optical devices.
\newblock {\em Microelectron. Eng.}, 220(15):111146, 2020.

\bibitem{intaravanne2020recent}
Yuttana Intaravanne and Xianzhong Chen.
\newblock Recent advances in optical metasurfaces for polarization detection
  and engineered polarization profiles.
\newblock {\em Nanophotonics}, 9:1003--1014, 2020.

\bibitem{huang2018graphene}
Zhong Huang, Kan Yao, Guangxu Su, Wei Ma, Lin Li, Yongmin Liu, Peng Zhan, and
  Zhenlin Wang.
\newblock Graphene-metal hybrid metamaterials for strong and tunable circular
  dichroism generation.
\newblock {\em Opt. Lett.}, 43(11):2636--2639, 2018.

\bibitem{zhou2019tunable}
Shaoen Zhou, Pengtao Lai, Hua Dong, Guo, Ping Li, Yuxiang Li, Zheng Zhu,
  Chunying Guan, and Jinhui Shi.
\newblock Tunable chiroptical response of graphene achiral metamaterials in
  mid-infrared regime.
\newblock {\em Opt. Express}, 27(11):15359--15367, 2019.

\bibitem{NPGAM2018_VO2Rev}
Zewei Shao, Xun Cao, Hongjie Luo, and Ping Jin.
\newblock Recent progress in the phase-transition mechanism and modulation of
  vanadium dioxide materials.
\newblock {\em NPG Asia Mater.}, 10:581--605, 2018.

\bibitem{falcone2004babinet}
F~Falcone, T~Lopetegi, MAG Laso, JD~Baena, J~Bonache, M~Beruete, R~Marqu{\'e}s,
  Ferran Mart{\'\i}n, and M~Sorolla.
\newblock Babinet principle applied to the design of metasurfaces and
  metamaterials.
\newblock {\em Phy. Rev. Lett.}, 93(19):197401, 2004.

\bibitem{zhu2012effect}
Yanhan Zhu, Yong Zhao, Mark Holtz, Zhaoyang Fan, and Ayrton~A Bernussi.
\newblock Effect of substrate orientation on terahertz optical transmission
  through {VO$_2$} thin films and application to functional antireflection
  coatings.
\newblock {\em J. Opt. Soc. Am. B}, 29(9):2373--2378, 2012.

\bibitem{bitzer2011terahertz}
Andreas Bitzer, Alex Ortner, Hannes Merbold, Thomas Feurer, and Markus Walther.
\newblock Terahertz near-field microscopy of complementary planar
  metamaterials: {B}abinet’s principle.
\newblock {\em Opt. Express}, 19(3):2537--2545, 2011.

\bibitem{OE2020_VO2AT}
Xiaoxiang Dong, Xiaoqing Luo, Yixuan Zhou, Yuanfu Lu, Fangrong Hu, Xinlong Xu,
  and Guangyuan Li.
\newblock Switchable broadband and wide-angular terahertz asymmetric
  transmission based on a hybrid metal-{VO$_2$} metasurface.
\newblock {\em Opt. Express}, 28:30675--30685, 2020.

\bibitem{yun2008vanadium}
Sun~Jin Yun, Jung~Wook Lim, JongSu Noh, ByungGyu Chae, and HyunTak Kim.
\newblock Vanadium dioxide films deposited on amorphous {SiO$_2$}-and
  {Al$_2$O$_3$}-coated {Si} substrates by reactive {RF}-magnetron sputter
  deposition.
\newblock {\em Jpn. J. Appl. Phys.}, 47(4S):3067, 2008.

\end{thebibliography}

\end{document}